\def\bfig{\begin{figure}}\def\fiz{{\hat F^i(z)}}\def\fizset{{\hat
F^{<i>}(z)}}  
\def\efig{\end{figure}}
\def\hk{\hat K}

\def\en{\hat E_N}

\def\gin{\gamma_i\cdot q([N])}
\def\fing{{\bf g}}
\def\gh{{\bf \hat g}}

\def\beq{\begin{equation}}
\def\eeq{\end{equation}}

\def\beqa{\begin{eqnarray}}
\def\eeqa{\end{eqnarray}}
\def\gp{{\bf G}}

\documentstyle[11pt,amssymbols,a4]{article}

\let\ssection=\section

\renewcommand{\section}{\setcounter{equation}{0}\ssection}
\begin{document}
\begin{titlepage}

\hfill Imperial/TP/92-93/38

\hfill {\tt hep-th 9306031}

\hfill {June 1993}

\vskip 3cm
\begin{center}
\huge
On the Topological Charges of Affine Toda Solitons
\vskip 2cm
\Large
Jonathan Underwood
\vskip 2cm\large
{\it Blackett Laboratory, Imperial College, London SW7 2AZ, United Kingdom.}
\normalsize
\vskip 2cm
\end{center}
\begin{abstract}
We provide a proof of a formula conjectured in \cite{OU93} for some
coefficients relevant in the principal vertex operator construction
of a simply-laced affine algebra $\gh$. These coefficients are important
for the study of the topological charges of the solitons of affine
Toda theories, and the construction of representations of
non-simply-laced $\gh$ and their associated Toda solitons.
\end{abstract}

\end{titlepage}

\bibliographystyle{unsrt}
\section{Introduction}

In this letter we provide a simple proof of a conjectured formula of
\cite{OU93}, whose notation we will be following, for some phases
$\epsilon(\tau_j,i)$ which appear in the description of the
principal vertex operator construction \cite{KKLW81} of the level one
representations of a simply-laced affine Lie algebra $\gh$. This can
be thought of as  an affinisation of the finite algebra $\fing$. We will
explain
the significance of the $\tau_j$ and $i$ below.

These phases have
several uses \cite{OU93}: \begin{itemize} \item They appear in the
formula for the vertex operator representation of the Kac-Moody
algebra valued $\fiz$, thus \beqa
\lefteqn{\rho_j(\fiz)=\epsilon(\tau_j^{-1},i)\exp\left(\sum_{N>0}{\gin z^N\hat
E_{-N}\over N}\right)}\nonumber\\
&&\hspace{5cm}\times\exp\left(\sum_{N>0}{-\gin^*z^{-N}\en\over
N}\right),\label{minreps}\eeqa where the $N$ run over the set of
positive exponents (see e.g. chapter 14 of \cite{Kac90}) of $\gh$.
These representations are important for the calculation of the
soliton solutions of an affine Toda theory. In \cite{OU93} it was
shown that these are generated by choosing a constant element of the
Kac-Moody group which appears in a specialisation of the
Leznov-Saveliev \cite{LS92} general solution of these theories to be
a product of exponentials of the form $\exp(Q\fiz)$. The index $i$
is a positive integer $\leq r$, where $r$ is the rank of $\fing$. It
labels the species of the soliton generated by the exponential.
\item In \cite{OU93} it was shown that the principal vertex operator
construction can in fact be used to evaluate the solitons solutions
of affine Toda theories with non-simply-laced $\gh$, by identifying
such $\gh$ as the subalgebras of some simply-laced algebras upon which
some outer automorphism acts trivially. We form the $\fizset$ for
this subalgebra as linear combinations of the original $\fiz$. To
determine the precise combination of these we need to know the expansion of
$\fiz$ over the generators of the Cartan subalgebra of $\gh$, and to
do this we need to know the $\epsilon$. \item They are the
characters of the irreducible one-dimensional representations of the
abelian group $W_0$, which is the subgroup of the diagram
automorphisms of the Dynkin diagram of $\gh$ which become inner
automorphisms when projected to automorphisms of the finite algebra
$\fing$. We shall describe $W_0$
further below. \item Solitons of an affine Toda theory can be
thought of as the solutions of least energy which interpolate the
degenerate vacua of the theory. The possible such vacua belong to
the co-weight lattice $\Lambda_W^*$ of $\fing$, and the difference between the
solution at $x=-\infty$ and $x=\infty$, known as the topological
charge, lies in this lattice. It has been a problem since the
original discovery of solitons solutions by Hollowood \cite{Ho92} to
determine these topological charges. It turns out \cite{OU93} that
the $\epsilon(\tau_j,i)$ associated to an $\fiz$ tell us which coset
of $\Lambda_W^*$ by the co-root lattice $\Lambda_R^*$ the topological
charge of the one soliton solution generated by $\exp(Q\fiz)$
belongs to.
\end{itemize}
\subsection{Definitions and Notation}
We shall now proceed to the derivation of a formula for the
$\epsilon$, for which we require some notation. Let us fix a Cartan
subalgebra of $\fing$, which we shall call $H'$. There is an element
$T_3$ of $H'$ whose adjoint action grades the step-operators of
$\fing$ corresponding to particular roots by the height of those
roots. This is the principal gradation. We may convert it to a
multiplicative gradation of the algebra using the Adjoint action of
the element $S$ defined by \beq S=\exp(2\pi i T_3/h).\label{sedff}\eeq
Here $h$ is the Coxeter number of $\fing$ which can be defined to be
one more than the height of highest root of $\fing$. It can be shown
(see e.g. \cite{FLO91}) that ${\rm Ad S}$ is actually the inner
automorphism of $\fing$ corresponding to a Coxeter element $w_C$
of the Weyl group, defined with respect to a second Cartan
subalgebra $H_0$. Such a subalgebra is said \cite{Ko59} to be in
apposition, and $H_0\cap H'=0$. Following \cite{FLO91} we bicolour the points
of the
Dynkin diagram of $\fing$ alternately black and white so that no
points of like colour are adjacent. Let us denote a product of Weyl
reflections in `black' simple roots by $w_B$ and $w_W$ as a product
of reflections in the `white' roots. The value of this, \cite{FLO91} is that
\beq w_C=w_Bw_W.\label{coxel}\eeq For later use let us define
$c(i)=\pm 1$, as $i$ is white or black, and $\delta_{iB}=1$ if $i$
is black and zero otherwise, with the complementary definition for
$\delta_{iW}$.

When the Dynkin diagram of $\gh$ has a symmetry $\tau$ there is a
natural lift of this to an outer automorphism of $\gh$, which can be
projected onto $\fing$ by setting the loop parameter to one for
example. Of these automorphisms some will become inner when
projected. This subgroup we denote by $W_0$. We now give some results proved
in \cite{OU93}.
\begin{itemize}
\item
$|W_0|$ is equal to the number of points $j$ on the Dynkin
diagram of $\gh$ which can be related to the point $0$ by a symmetry
$\tau$.
\item
For each such point $j$ there is exactly one $\tau_j\in W_0$ such
that $\tau(0)=j$. This serves to label the elements of $W_0$.
\item
Let us denote the realisation of $\tau_j$ as an inner automorphism
of $\fing$ by ${\rm Ad}T_j$. Then
\beq T_j S T_j^{-1}=S\exp\left(-2\pi i {2\lambda_{\tau(0)}\cdot H'
\over\alpha_{\tau(0)^2}} \right).\label{sts}\eeq
\item
Ad$T_j$ acts trivially on $H_0$ and so $T_j\in\exp H_0$.
\end{itemize}

\section{Calculation of the phases $\epsilon$}\label{epscalc}

We want to calculate directly the action of the $\tau\in W_0$ on the
$\fiz$. Because the $\tau_j$ can be realised as inner automorphisms
${\rm Ad} T_j$ in $\fing$ we can determine this action if we
explicitly solve for $T_j$. We already know that $T_j\in\exp H_0$
and so writing $T_j=\exp (2\pi i Y_j\cdot H_0)$ and using the
definition of the $\epsilon$ \beq T_j
E_{\gamma_k} T_j^{-1}= \epsilon(\tau_j,k)
E_{\gamma_k}.\label{tconj}\eeq we find that \beq
\epsilon(\tau_j,k)=e^{2\pi i Y\cdot\gamma_k}.\label{phases1}\eeq
$E_{\gamma_k}$ are the step-operators with respect to $H_0$, which
yield $\hat F^k(z)$ when lifted to $\gh$ in an appropriate fashion.
It is sufficient to use only $\gamma_k=c(k)\alpha_k$ to obtain $\hat
F^k(z)$ whose modes span $\gh$ (\cite{FLO91}, \cite{OTU93}).

Rearranging equation \ref{sts} yields \beqa ST_jS^{-1}&=&Te^{2\pi
i(2\lambda_j\cdot H'/\alpha_j^2)}\nonumber \\ &=&Te^{2\pi
i(2\lambda_j\cdot H_0/\alpha_j^2)}.\label{tst}\eeqa It is important
to note the small but significant difference between the right-hand
sides of this expression. In one case the central element is
expanded over $H'$, and the second over $H_0$. The equality of these
expressions follows precisely because the element is central, so
that any conjugation of $\gp$ sending $\exp H'$ to $\exp H_0$ will
leave the central elements unchanged.  We can then use
\ref{tst} to find \beq \exp \left(2\pi i
\left(w_C-1\right)(Y_j)\right) = \exp\left( 2\pi
i\frac{2\lambda_j\cdot H_0}{\alpha_j^2}\right),\label{yform}\eeq and
so solving this expression we find that \beq \left(w_C-1\right)Y_j\in
\frac{2\lambda_j}{\alpha_j^2} +\Lambda^*_R,\label{y2}\eeq yielding
\beq Y_j=\left(w_C-1\right)^{-1}\frac{2\lambda_j}{\alpha_j^2} \bmod
\Lambda_R^*.\label{yfin}\eeq Note that $w_C-1$ is invertible since
none of the eigenvalues of $w$ is unity.

Let us examine explicitly the action of $(w_C-1)$ on the basis of
$H_0^*$ provided by the fundamental weights of $\fing$. Let us drop
the subscript. We find,
using \ref{coxel},
\beqa \left(w-1\right)\lambda_i&=&-\alpha_i,\quad i\quad{\rm black};\nonumber
\\ =-w\left(w^{-1}-1\right)\lambda_i&=&w  (\alpha_i),\quad i\quad{\rm
white}.\eeqa Inverting these expressions produces \beq
\left(w-1\right)^{-1}\alpha_i=\left\lbrace \begin{array}{l} -\lambda_i,\quad
i\quad {\rm black};\\ \lambda_i-\alpha_i,\quad i\quad {\rm
white}.\end{array}\right. \label{wmoinver}\eeq To find the action of
$(w-1)^{-1}$ on $\lambda_j$, we need to expand it over the simple
roots, as \ref{wmoinver} gives us the images of these.
 The relevant expansion can easily be checked to be \beq
\lambda_j=K_{ji}^{-1}\alpha_i,\label{lamexp}\eeq and so substituting
in we get \beq (w-1)^{-1}\lambda_j=-\sum_{i\quad {\rm
black}}K_{ji}^{-1}\lambda_i +\sum_{i\quad {\rm
white}}K_{ji}^{-1}(\lambda_i-\alpha_i).\label{junk}\eeq Defining the traceless
Cartan matrix $\hk_{ji}=K_{ji}-2\delta_{ji}$ we can then rearrange
the white sum to
make the above look more symmetrical \beqa \sum_{i\quad {\rm
white}}K_{ji}^{-1}(\lambda_i-\alpha_i)&=&\sum_{i\quad {\rm
white}}K_{ji}^{-1}\left(-\lambda_i-\hk_{il}\lambda_l\right)\nonumber
\\ &=& -\sum_{i\quad {\rm
white}}K_{ji}^{-1}\lambda_i-
\sum_{i,l}K_{ji}^{-1}\hk_{il}\lambda_l\delta_{lB}\nonumber \\&=&-\sum_{i\quad
{\rm
white}}K_{ji}^{-1}\lambda_i-
\sum_{i,l}K_{ji}^{-1}(K_{ij}-2\delta_{il})\lambda_l\delta_{lB}\nonumber
 \\&=&-\delta_{jB}\lambda_j+\sum_i
K_{ji}^{-1}(2\delta_{iB}-\delta_{iW})\lambda_i.\label{longtedious}\eeqa
In this calculation the crucial point is that $\hk_{ij}$ contains
only cross terms between white and black indices, and so we are able
to replace a sum over purely white indices by a sum over all of
them. Now we can rewrite \ref{junk} in the form \beq
(w-1)^{-1}\lambda_j=-\delta_{jB}\lambda_j+\sum_i
K_{ji}^{-1}c(i)\lambda_i.\label{temp1}\eeq

Using \ref{temp1} and \ref{yfin} in \ref{phases1} we find that \beq
\epsilon(\tau_j,k)=\exp\left(2\pi i\frac{K_{\tau
k}^{-1}\alpha_k^2}{\alpha_\tau^2}\right).\eeq Substituting  $K_{\tau
k}^{-1}=2\lambda_\tau\cdot\lambda_k/\alpha_k^2$ finally yields \beq
\epsilon(\tau_j,k)=\exp\left(2\pi
i\frac{2\lambda_\tau\cdot\lambda_k} {\alpha_\tau^2}\right),\label{epsfin}\eeq
which proves the conjecture of \cite{OU93}, in a corrected
form.\footnote{The reason the sign of the exponential differs is due
to an incorrect assignment of eigenvectors in \cite{OU93}.}

\section{Discussion}
It is important to have a proof of the formula \ref{epsfin}, as it
has since been used by Kneipp and Olive \cite{KO93} to derive an interesting
identity in a given representation of the Kac-Moody group which encapsulates
the crossing of
soliton into antisoliton. Of course \ref{epsfin} was also important
for the work of \cite{OU93}, where it appeared to rationalise a
number of things known at the time.

It would be interesting to know if this sort of argument could be
applied to the more general vertex operator constructions described
in \cite{KP85} and used to construct non-abelian Toda theories and
their soliton-like solutions in \cite{Un93}.

Finally we note that a variety of interesting identities for the
Coxeter element and its eigenvectors are already known (see for
example \cite{Do91}, \cite{Do92}, \cite{FO92}).

\subsection{Acknowledgements}
I would like to thank David Olive for valuable discussions on the
material mentioned in this letter. I would also like to thank the
United Kingdom Science and Engineering Research Council for the
funding under which this research was carried out.

\end{document}